\documentclass{article}

\usepackage{arxiv}
\usepackage[T1]{fontenc}
\usepackage{graphicx}
\usepackage{amsmath}
\usepackage{algorithmicx}
\usepackage{graphicx}
\usepackage{xcolor}
\usepackage{booktabs}
\usepackage{multirow}
\usepackage{colortbl}
\usepackage{array}
\usepackage{dashrule}
\usepackage{placeins}
\usepackage{hyperref}

\newcommand{\sln}{ConMover}

\title{\sln{}: Generating Move Smart Contracts based on Concepts}

\author{\href{https://orcid.org/0000-0002-6705-6506}{\includegraphics[scale=0.06]{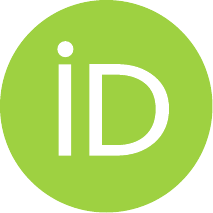}\hspace{1mm}Rabimba Karanjai}\footnotemark[1] \\
	Department of Computer Science\\
	University of Houston\\
	Houston, TX, United States \\
	\texttt{rabimba@cs.uh.edu} \\
        \And
        Sam Blackshear \\
        Mysten Labs \\
        \And
	Lei Xu \\
	Department of Computer Science\\
	Kent State University\\
	Kent, OH, United States \\
	\And
	Weidong Shi \\
	Department of Computer Science\\
	University of Houston\\
	Houston, TX, United States \\
}
\begin{document}

\maketitle              % typeset the header of the contribution
\begin{abstract}
The increasing use of formal verification for smart contracts has led to the rise of new, formally verifiable languages, such as Move. However, the scarcity of training data for these languages poses a challenge to code generation using large language models (LLMs). This paper introduces ConMover, a framework that utilizes a knowledge graph of Move concepts and a small set of correct Move code examples to enhance the code generation capabilities of LLMs. ConMover employs a novel approach that combines concept retrieval, planning, coding, and debugging agents to iteratively refine the generated code. This framework is evaluated with different sized open-source LLMs, demonstrating significant improvements in code generation accuracy compared to baseline models. The results highlight the potential of ConMover to bridge the gap between natural language descriptions and low-resource code generation for smart contracts, enabling more efficient and reliable development processes.

\end{abstract}
\section{Introduction}
The growing significance of code generation in automating software development has spurred extensive research into utilizing Large Language Models (LLMs)~\cite{li2022competition,chen2021evaluating} for this purpose. Although LLMs have shown impressive skill in translating natural language instructions into code, they still struggle to produce flawless code for complex tasks on the first attempt~\cite{zheng2023codegeex,wang2023codet5+}. This challenge mimics the experience of human developers, who often go through several cycles of debugging and refinement for intricate coding problems. As a result, the focus is increasingly shifting towards enhancing LLMs' ability to self-debug—detecting and fixing errors in their own code—thereby strengthening their overall code generation performance.

The limitations of single-pass code generation are becoming increasingly apparent. Generating code in a single attempt often fails to capture the nuances and intricate requirements inherent in bigger code repositories~\cite{jiang2023self}. This approach struggles to address the numerous edge cases and specific needs that arise, particularly given the high degree of precision and complexity demanded in such projects. 
To address these shortcomings, researchers are transitioning towards multi-round code generation frameworks~\cite{nijkamp2022codegen} that leverage iterative refinement. These frameworks facilitate a more robust and accurate code generation process by enabling the model to generate and refine code through multiple iterations. This approach allows for incremental improvements and adjustments, ultimately leading to a more accurate and efficient development process. These advanced systems employ reflective mechanisms, such as analyzing failed test cases and interpreting error messages, to inform subsequent code generation attempts~\cite{chen2023teaching}. While these techniques have demonstrated promising improvements, they still produce considerably worse code when it comes to complex tasks~\cite{karanjai2024harnessing}.

Moreover, when it comes to generating code for low-resource languages like Move\cite{blackshear2019move} or Rust, prior work \cite{eniser2024towards,karanjai2024solmover} has shown an approach of translation works best. However, that might not be sufficient for most of the use cases, and a method that does not rely on existing code in a different programming language or vast code corpora to train a model for these low-resource languages is needed. 

We try to address the following research questions through this work.

\begin{itemize}
    \item \textbf{RQ1:} Can we generate relevant concepts based on code documentation and coding guides that are useful for an LLM for code generation
    \item \textbf{Rq2:} Can we use a very small amount of code and use them as schemas to generate code in that language without prior finetuning on that coding language
    \item \textbf{RQ3:} Is it possible to use \sln{} like framework to augment smaller LLMs to produce better code.
    \item \textbf{RQ4:} Can we provide a self-correction method that utilizes concepts instead of existing code to correct the initial generated code candidate
\end{itemize}

In this paper, we propose \sln{}, a framework that can generate code without pre-training an LLM on a vast amount of existing code corpora based on code definition, concepts, and rules. It also refines and corrects its generated code for better quality. We first go into details of generating coding concepts and rules for Move. They will act as our evaluators for final code refinement. We also go into details of how, with a small sample size of existing Move code, we are able to guide the model for correct Move code generation. Supervised fine-tuning (SFT) on high-quality demonstration data significantly enhances the ability of LLMs to explain incorrect code and refine it, outperforming existing prompting techniques by a substantial margin. Additionally, we introduce a reinforcement learning (RL) approach with an innovative reward system that accounts for both the semantics of code explanations and the success rate of unit tests. This enables LLMs to generate more effective code explanations and make accurate refinements. 

This paper makes the following contributions:
\begin{itemize}
    \item We introduce \textbf{ConMover}, a framework for generating Move code using LLMs by leveraging a knowledge graph of Move concepts and a small set of correct code examples.
    \item ConMover employs a multi-agent approach, integrating concept retrieval, planning, coding, and debugging agents to refine the generated code iteratively.
    \item We conduct experiments with various open-source LLMs, demonstrating that ConMover significantly improves code generation accuracy, particularly for smaller models.
    \item Our findings highlight the potential of ConMover to enable efficient and reliable smart contract development by facilitating code generation from natural language descriptions for low-resource languages like Move.
\end{itemize}

\section{Approach}

% We want to generate smart contracts based on user given tasks. And one of our candidate languages for smart contract generation, Move is a low resource language. They don't have enough source code available for fine tuning or pretraining a LLM. That is one of the significant challenges we try to tackle in this paper.
To generate smart contracts from user tasks, we chose Move as our programming language. However, limited Move resources pose a challenge for training effective code generation models. Our approach addresses this by using Move documentation to create a knowledge graph and an RAG-based validator. This validator acts as a judge, ensuring the logical correctness of the generated code.

% In this section we detail our framework, the different parts of the Concept generation, Code generation as well as code correction mechanism that has been deployed. We detail our different parts of the architecture, high level design and the reasoning behind them.
In subsequent subsections, we will discuss in detail the different stages in our proposed framework, which includes Concept generation, Code generation as well as Code correction mechanism. 
We would also be discussing in detail the architecture, high-level design, and the reasoning behind them. 

\subsection{Generating Concepts} \label{genconsept}
In this section, we define how we process the Move documentation data. As a data source we intentionally only use \cite{MoveConceptsSuiDocumentation-2024-10-11} to create our knowledge base. For knowledge base creation we use existing knowledge graph techniques, or more specifically Docs2KG~\cite{sun2024docs2kg} to create our knowledge graph.

When a LLM is asked to write a Move code without any modification. It produces some code that looks like move that might or might not compile. We have seen from \cite{karanjai2024solmover} that these codes primarily end up being wrong rust codes since move is based on rust. In our experiment, we have also seen it produces move code that is a mix of move code compatible with Sui and apros chain. For our use case we want it to produce code correct for one specific chain based on user input. We have mostly experimented with Sui blockchain for our work.

Instead of forcing the LLM to generate a formal structured query (like SQL) to interact with the Knowledge Graph database, this approach takes a more intuitive route. The LLM is fine-tuned to produce a natural language query that describes the LLM-generated Move Code. This offers several advantages. Firstly, it promotes natural language queries that tend to be more succinct. Secondly, it avoids the impracticality and potential performance issues associated with the training of LLM on millions of variable IDs that can appear if the Knowledge Graph gets bigger.
Finally, expressing the query in natural language is a simpler task for the LLM, as to rephrasing and extracting information from the surrounding context. This approach essentially streamlines the process of connecting LLM-generated code with relevant data in our knowledge base, making concept generation better.

To achieve the desired behavior, the model is fine-tuned using an instruction-response dataset. This approach parallels techniques utilized in tool use for LLMs, emphasizing the adaptation of the model to effectively leverage external tools rather than depending solely on next-token prediction for generating responses\cite {schick2024toolformer}. The objective is to ensure that the fine-tuned model maintains the original model's fluency and natural language generation style, similar to what is observed in the tool-use domain. 

In our implementation, the answer is presented alongside the original LLM-generated answer as a way for the Code Generator to use these as context.

\subsection{Data collection and verification}

 We first try to collect incomplete and wrong codes generated by a stock LLM. We use \cite{electriccapital} as our coding data source to generate task descriptions for the LLMs to solve.  We use 313 code examples from the whole dataset to create our wrong solution dataset. To augment this we prompt Gemini 1.5 with 2 shot examples to sample 20 solutions for each of those task descriptions. Once we have all the generated candidates, we run it through the compiler and collect all the execution traces and errors.

\section{\sln{}}

This research introduces \sln{}, a multi-agent framework designed to tackle generation of Move Smart Contracts through a collaborative code generation process. Mimicking the human approach to programming, \sln{} employs specialized Large Language Model (LLM) agents: retrieval, plan, code, and debug. These agents operate in a structured pipeline, with each agent's capabilities enhanced by leveraging the outputs of preceding agents. However, recognizing that not all agent outputs are equally valuable, \sln{} incorporates an adaptive traversal scheme that allows dynamic interaction between agents. This enables iterative code improvement, such as fixing bugs, while optimizing the utilization of each LLM agent. This section goes into the details of each agent, their specific roles, prompt structures, and how they interact within the MapCoder framework to effectively generate solutions for competitive programming challenges.

\subsection{Concept Retreival Agent}

The Retrieval Agent in \sln{} functions like a memory bank, drawing on past experiences to address new challenges. Rather than relying on manual input or separate retrieval models, this agent uses the output from our concept generator as a guidance scheme.

The concept generator uses the Move Code book and sui examples to produce solution code step-by-step; the agent can extract sequences of thought that help it formulate corresponding plans. Additionally, it is tasked with generating relevant algorithms and tutorials, encouraging reflection on the underlying algorithmic principles and enabling the creation of algorithmically related examples\cite{yasunaga2022retrieval}. This holistic approach ensures that the Retrieval Agent delivers rich and contextually relevant information, supporting the code generation pipeline in its subsequent stages.

\subsection{Planner Agent}

The second agent in the \sln{} framework, known as the Planning Agent, is responsible for creating a detailed, step-by-step plan to solve the given programming problem. It builds on the examples and their corresponding plans retrieved by the Retrieval Agent.

Rather than merging all examples into a single plan, the Planning Agent takes a more refined approach. It generates individual plans for each example, recognizing that not all retrieved examples are equally relevant. This method is based on the insight that the order and relevance of in-context information greatly influence the LLM’s reasoning capabilities~\cite{xu2023re}. By generating multiple plans, the agent creates diverse problem-solving pathways, offering flexibility in approach.

To improve this process further, the Planning Agent is designed to not only produce plans but also evaluate their confidence levels. This is done by prompting the LLM to assign a confidence score to each plan. These confidence scores help guide the subsequent agents in the pipeline, allowing them to prioritize and select the most promising plans for code generation. This ensures that the Planning Agent maximizes the value of the retrieved information while providing critical insights for the later stages of the code generation process.

\subsection{Coding Agent}

The Coding Agent in \sln{} plays a pivotal role in the code generation process. It takes the problem description and the selected plan from the Planning Agent and translates the plan into executable code.

As part of the pipeline, the Coding Agent receives both the original problem and a specific plan, which it uses to generate the corresponding code. Once the code is generated, it is immediately tested against predefined sample inputs and outputs. If the code passes these tests, it is deemed a potential solution. However, if the code fails, it is forwarded to the next stage, the Debug Agent, for refinement and correction. This iterative process ensures that the generated code is continuously validated and enhanced, improving the chances of arriving at a correct and efficient solution.

\subsection{Debugging and feedback Agent}

The Debugging Agent serves as the final safeguard in \sln{}’s code generation pipeline, with its main responsibility being to identify and resolve errors in the code produced by the Coding Agent.

Mirroring the approach of human developers who often refer back to their original plan during debugging, the Debugging Agent is also equipped with the plan from the Planning Agent. This "plan-aware" debugging approach significantly enhances the agent's ability to detect and correct bugs, underscoring the important relationship between planning and debugging in the overall code generation process.

A key advantage of the Debugging Agent is that it works directly with the sample input/output provided in the problem description, eliminating the need for additional test case generation. This not only streamlines the debugging process but also sets \sln{} apart from other systems that require separate test case creation steps.

For each plan generated by the Planning Agent, the Debugging Agent refines the code iteratively, ensuring thorough error correction. This combination of iterative refinement and plan-aware debugging allows \sln{} to address issues and produce high-quality, reliable code solutions efficiently.

\subsection{Dynamic Traversal}

\sln{} employs a dynamic and iterative process for code generation, closely resembling the workflow of a human programmer. The process starts with the Planning Agent creating multiple plans, each assigned a confidence score. These plans are ranked, and the Coding Agent is tasked with translating the highest-scoring plan into executable code. The generated code is then tested against sample inputs and outputs. If it passes the tests, it is returned as a solution. If the code fails, the Debugging Agent takes over, iteratively refining the code to fix any detected errors.

The debugging process continues for a predefined number of iterations. If the code remains unsuccessful after these attempts, the system loops back to the Planning Agent, which selects the next highest-confidence plan. This cycle of planning, coding, and debugging repeats until a correct solution is found or the maximum number of iterations is reached.

This dynamic agent traversal strategy allows \sln{} to explore different problem-solving approaches efficiently while continuously refining the generated code. By iterating through various plans and corrections, the system significantly increases the likelihood of producing a correct and reliable solution.

\section{Architecture of ConMover}

CoMover adopts an agentic approach of holistically generating move code.
Each agent has multiple parts inside it not only a LLM. 

\subsection{Planner Agent}

\begin{figure}
    \centering
    \includegraphics[width=0.8\linewidth]{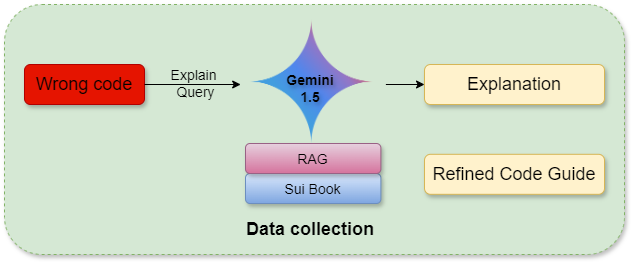}
    \caption{Concept Generation}
    \label{fig:Concept Generation}
\end{figure}

The planner agent encapsulated the concept generator inside it. Its workflow looks Figure \ref{fig:Concept Generation}.

Here We take the faulty codes that we intentionally generated by prompting the LLMs as well as initial fault codes that acted as seed to build our wrong code corpora. We run it through the compiler and log all the execution trace and compilation errors. These will act as our fault context. These are then passed to a more capable LLM, Gemini 1.5 (V 002). gemini here is augmented by a RAG stack. This stack supplies contextual information based on the Sui Move documentation as described in Section \ref{genconsept}.

We divide our task in the following parts. \((Q, T_v, T_h, E)\). Here, \(Q\) represents the task description, which includes code snippets and requirements in natural language. \(T_v\) stands for visible test cases, \(T_h\) for hidden test cases, and \(E\) is the entry point. All test cases are executed starting from the entry point.
\section{Reasoning and Memory Management}

This plays a key role in orchestrating initial reasoning and strategic thinking. This agent dynamically adapts its input information based on the task's inherent complexity and the current execution stage. Moreover, the concept generation is tasked with selecting and integrating pertinent guidance retrieved from a dedicated Memory Pool. However, to optimize token utilization and mitigate potential LLM overhead from excessive context, memory pool lookups and extractions are strategically bypassed during the initial code generation phase using the KnowlegdeGraph as data store.
\begin{figure}
    \centering
    \includegraphics[width=1\linewidth]{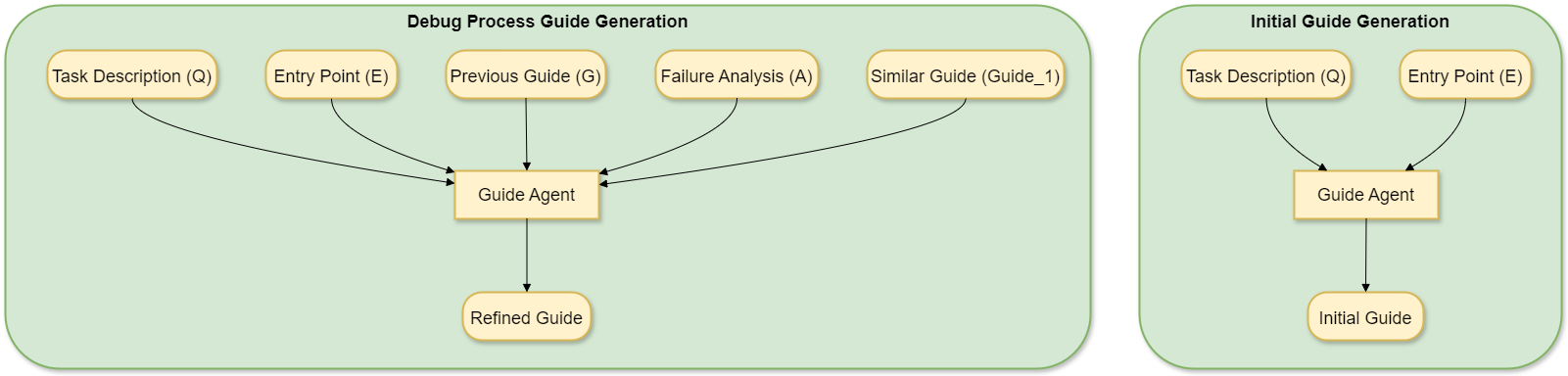}
    \caption{Planning Agent}
    \label{fig:plan-label}
\end{figure}
Figure \ref{fig:plan-label} illustrates how distinct prompts are employed during the initial and debug stages. In the initial stage, the Planning Agent generates a Generation plan based exclusively on the task description ($Q$) and entry point ($E$), without any additional information. During the debug stage, however, the system matches and retrieves relevant samples from the memory pool based on task similarity. The matching mechanism considers the following key components:

$Q$: The provided task description.

$P$: The plan generated by the Planning Agent.

$K$: A set of keywords extracted from the task description ($Q$) and the generated code ($C$).

$E$: The execution results, which encompass both visible test cases ($T_v$) and hidden test cases ($T_h$).

The Memory Pool ($\mathcal{M}$) stores tuples structured as $(Q_i, P_i, K_i)$, where:

$Q_i$ represents a previously encountered task description.

$P_i$ denotes the associated generation guide for that task.

$K_i$ signifies the set of keywords extracted using the GPT-4o-mini API after successful task completion (passing both $T_v$ and $T_h$).

When a task is successfully completed (passing all test cases), the Gemini API is called upon to extract a set of keywords ($K_i$) based on the task description ($Q$) and the generated code ($C$). This extraction process is formalized as:

\begin{equation}
K_i = \text{ExtractKeywords}(Q, C)
\end{equation}

Here, $\text{ExtractKeywords}$ is the function responsible for extracting relevant keywords from the task description and its corresponding solution. When a new task ($Q_{\text{new}}$) is encountered during the Debug phase, the memory pool is queried for similar tasks. 

% This retrieval process employs a hybrid strategy, using SBERT \cite{reimers2019sentence} \cite{feng2020codebert} for description matching and BM25 \cite{lin2021few} for index term matching, ultimately achieving a hybrid retrieval system \cite{karpukhin2020dense}. Algorithm \ref{alg:memory_pool} presents the pseudo-code for the memory pool matching functions, where the $\text{Sim}$ function calculates the similarity between $Q_{\text{new}}$ and $Q_i$, as well as between their respective keywords.

% The Planner LLM then constructs a final plan ($P_{\text{final}}$) by augmenting the initial plan ($P_{\text{init}}$), generated from $Q_{\text{new}}$, with the most relevant plan ($P_i$) retrieved from the memory pool based on the similarity score:

% \begin{equation}
% P_{\text{final}} = \text{Planner}({P_{\text{init}}, P_i, Q, A})
% \end{equation}

% where $A$ represents the failure analysis provided by the Feedback Agent.

\subsection{Coding Agent}

The coding agent is a plug-and-play module in our framework. This can take any pre-trained LLM and generate code based on the given input. We have intentionally refrained from any fine-tuning steps in the Coding Agent LLM. That makes our approach universally applicable for any LLM and can be benchmarked to determine comparative performance.

\begin{figure}
    \centering
    \includegraphics[width=0.8\linewidth]{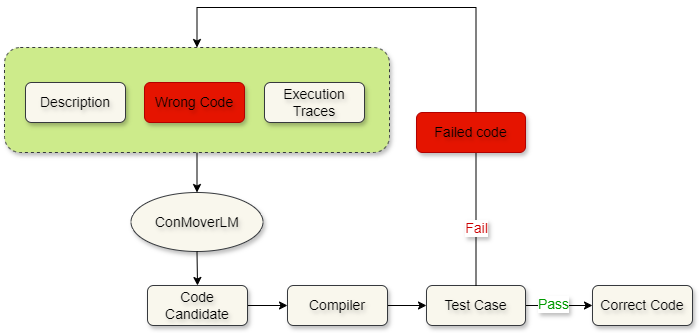}
    \caption{Feedback Loop}
    \label{fig:fdloop}
\end{figure}

\subsection{Debugging and Code Refining}

The debugging and feedback agent depends on the planner and knowledge graph to have feedback for the generated code. Code Language Models \cite{li2022competition,zheng2023codegeex,roziere2023code,wang2023codet5+}(LLMs) are typically trained on massive datasets of source code, learning to predict sequences of code tokens. However, the inherent noise within these datasets, often sourced from open-source repositories~\cite{kocetkov2022stack,roziere2023code,xu2022systematic}, can impact the accuracy and reliability of the generated code.

The quality of code in these datasets varies significantly. High-quality code, often from well-maintained projects, aligns well with its documentation, while noisy code, potentially from less experienced developers or unfinished projects, may contain vulnerabilities or inconsistencies. This means LLMs are exposed to both correct and incorrect code during training, potentially learning and replicating undesirable patterns~\cite{li2022competition}.

Consequently, when prompted to generate code, these LLMs might produce either accurate or erroneous outputs, influenced by the ratio of correct to incorrect code encountered during training. This inherent unpredictability can lead to unexpected behaviors and even security vulnerabilities in the generated code~\cite{he2023large}.

To mitigate the potential for code LLMs to generate erroneous or harmful code during knowledge distillation, a focused fine-tuning step is introduced. This involves training the LLMs exclusively on "guaranteed correct" code, specifically canonical solutions to programming challenges validated by test suites.

Unlike traditional pre-training that predicts every token in the code corpora, this fine-tuning process concentrates solely on the code tokens within these correct solutions. The accompanying natural language descriptions are treated as context, guiding the model towards understanding the intent and functionality of the code.

More formally, given a natural language description $NL = \{nl_0, nl_1, \dots, nl_m\}$ with $m$ tokens and a canonical solution $C = \{c_0, c_1, \dots, c_n\}$ with $n$ tokens, the fine-tuning process utilizes the standard language modeling loss to optimize the model:

\[
\mathcal{L}_{\text{fine-tune}} = \sum_{n \in |C|} -\log P(c_n \mid NL, c_1, c_2, \dots, c_{n-1})
\]

This targeted fine-tuning approach aims to refine the LLM's code generation capabilities by exposing it to correct code examples exclusively. By learning from these verified solutions and leveraging the contextual information provided by natural language descriptions, the model is trained to generate more accurate and reliable code. This preemptive step helps minimize the risk of unexpected or malicious behaviors during subsequent knowledge distillation and self-refinement stages.

To effectively train code language models (LMs) in code refinement, we propose a template that consolidates data from various sources. 
This template, illustrated in Figure~\ref{fig:template}, is designed to preserve the naturalness of code by embedding it within docstrings or comments. It consists of the following key elements:

\begin{figure}
    \centering
    \includegraphics[width=1\linewidth]{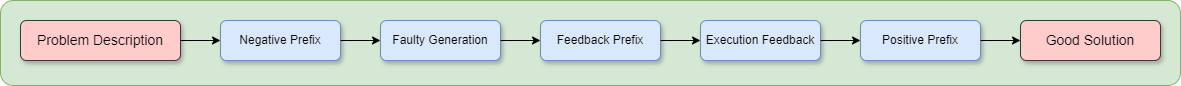}
    \caption{Template}
    \label{fig:template}
\end{figure}

By combining these components, the template provides a structured method for guiding the model from flawed code to a corrected solution.

The proposed template effectively gathers all the necessary information for the code LLM to learn and perform code refinement. This aggregated information, denoted as  AGGR={NL,FG,EF}, includes the natural language problem description (NL), the faulty generated code (FG), and the execution feedback (EF). The model is then trained to predict the canonical solution (C) token by token,  guided by this aggregated context.

$\mathcal{L}_{\text{self-refine}} =  \sum_{n \in |C|} -\log p(c_n | NL, FG, EF, c_1, c_2, ..., c_{n-1})$

While the core of this learning process relies on next-token prediction, it proves surprisingly effective in teaching the code LLM self-refinement. This success stems from the learning objective compelling the model to integrate knowledge from diverse sources.

To accurately predict each code token $c_i$, the model must effectively understand the natural language problem description (NL), the faulty generated code (FG), and the execution feedback (EF). It then needs to synthesize this information, determining how to leverage each resource to generate the correct code. This multi-faceted understanding and decision-making process contributes significantly to the model's ability to self-refine and improve its code generation capabilities.

\subsection{Code Correction based on feedback}

\sln{} depends on compiler feedback as well as the knowledge graph in the planning stage to decide its correction phase. ConMoverLM is fine-tuned on faulty code execution traces and its compilation errors. The instruction tuning dataset we used for the training is based on Starchat \cite{star} instruction tuning dataset, but modified based on our compilation error dataset.

ConMoverLM creates the initial code candidate based on the initial code description. The initial code candidate goes through a compilation check. If it compiles, it goes through the test cases. If it fails the test case, it goes back to an iterative process, but this time, in the input string, the previous faulty code gets replaced by the new faulty code along with the failed execution trace supplied by the planner retrieved from the memory pool. This iterative process goes through five times and we log all the successful compilations.

\section{Experimental Setup}

\subsection{Model}

To ensure \sln{}'s generalizability, we evaluated our framework with three open-source LLMs, both with and without our proposed enhancements. We also compared its performance against state-of-the-art models. Specifically, we employed CodeLlama-7B~\cite{roziere2023code}, Gemma-2 9B, and Gemma-2 2B~\cite{team2024gemma}. The inclusion of a 2B parameter model allowed us to assess how our framework impacts the performance of smaller LLMs, which are generally perceived as less capable.

\subsection{Configurations and Hyperparameters}

We conducted our experiment on a machine with 4xNvidia A100 with 80GB memory.

 Training involves feeding the model 512 examples at a time, each containing up to 2,048 units of text (BPE tokens). The training process uses a common technique where the learning rate—how quickly the model adapts—starts high and gradually decreases. Smaller versions of \sln{} use higher learning rates than larger ones.

Specifically, they fine-tune the model first, training it on a prepared dataset for one epoch (a complete pass through the data) with learning rates varying from 5e-5 to 1e-5 depending on the model size.  Then, they employ a "self-refinement" phase, again training for one epoch, using learning rates from 2e-5 down to 5e-6.  A "cosine learning rate decay scheduler" with warmup steps is used throughout.

During self-refinement, 5\% of the input is masked out, and 25\% of the training data consists of self-refined samples.

When generating text with \sln{}, we use a method called "nucleus sampling" with a top-p probability of 0.95.  This helps ensure the generated text is both diverse and coherent.  The model is allowed to produce up to 256 BPE tokens for most tasks, but up to 512 tokens for more complex problems that require longer solutions.

Inference also includes self-refinement, which is limited to a maximum of 4 steps to balance improvement with speed. Further sections of the paper analyze the impact of varying the number of refinement steps on the overall inference speed.

\section{Evaluation And Results}

This section evaluates \sln{}'s code generation capabilities, comparing its performance to MOve for various sizes across our scraped dataset in Move\cite{electriccapital}. We assess \sln{} in two primary settings: one-time generation and iterative self-refinement.

\noindent\textbf{One-Time Generation.}  Following the established evaluation protocols of the benchmarks \cite{chen2021evaluating,austin2021program}, we provide the model with a description of the programming task as a prompt and evaluate the generated code against accompanying test suites. Each test suite comprises multiple test cases designed to assess the code from different perspectives.

\noindent\textbf{Iterative Self-Refinement.}  Mirroring \sln{}'s inference framework, we utilize the one-time generation output as the initial input for the self-refinement process. This process is repeated up to five times.  A successful refinement is achieved when the generated code passes all test cases within the allotted iterations. Conversely, if the model fails to produce a correct program within four refinement attempts, the sample is deemed a failure.

% \usepackage{graphicx}
\begin{table}[]
\resizebox{\columnwidth}{!}{%
\begin{tabular}{|l|cc|}
\hline
\textbf{Model}               & \multicolumn{1}{c|}{\textbf{One-time}} & \textbf{Self-Refine}        \\ \hline
                             & \multicolumn{2}{c|}{\textbf{Move Test (780 Tests)}}                  \\ \hline
Gemma2 2B                    & \multicolumn{1}{c|}{12.2}              & 12.2$\sim$$\sim$\{+0.0\%\}  \\ \hline
Gemma2 2B+ RAG w/ Correct    & \multicolumn{1}{c|}{12.2}              & 14.0$\sim$\{+14.9\%\}       \\ \hline
ConMover+ KG Planner         & \multicolumn{1}{c|}{14.0}              & 20.7$\sim$\{+47.9\%\}       \\ \hline
CodeLlama-7B                 & \multicolumn{1}{c|}{15.9}              & 16.5$\sim$$\sim$\{+3.8\%\}  \\ \hline
CodeLlama-7B+ RAG w/ Correct & \multicolumn{1}{c|}{18.3}              & 18.9$\sim$$\sim$\{+3.3\%\}  \\ \hline
ConMover+ KG Planner         & \multicolumn{1}{c|}{18.3}              & 22.0$\sim$$\sim$\{+20.0\%\} \\ \hline
Gemma2 9B                    & \multicolumn{1}{c|}{21.9}              & 23.8$\sim$$\sim$\{+8.4\%\}  \\ \hline
Gemma2 9B+ RAG w/ Correct    & \multicolumn{1}{c|}{21.9}              & 23.8$\sim$$\sim$\{+8.4\%\}  \\ \hline
ConMover+ KG Planner         & \multicolumn{1}{c|}{21.4}              & 29.3$\sim$$\sim$\{+37.1\%\} \\ \hline
\end{tabular}%
}
    \caption{Comparing \sln{}’s performance with baseline models in both one-time generation and iterative
self-refinement settings.}
    \label{tab:ref1}
\end{table}

Table \ref{tab:ref1} shows us that Self-Refinement Capacity Does Not Come Along Naturally with Code LMs’ Pre-training. Although large language models are trained on extensive code datasets, they are primarily designed for one-time code generation.  They struggle to improve their code by analyzing the results of execution and identifying errors.

Simply increasing model size doesn't improve self-correction abilities, indicating current training methods lack the necessary focus on self-refinement.  Augmenting models with correct code and knowledge through RAG shows only marginal improvements, highlighting that simply providing information is insufficient for  models to rectify their own mistakes. They need a deeper understanding of execution feedback and error correction.

Our approach, \sln{}, consistently improves correct Move code generation through self-refinement, achieving up to a 47.1\% relative improvement. Notably, while beneficial for all models, the greatest gains are observed with the smaller Gemma2 2B model, demonstrating the effectiveness of \sln{}'s self-refinement and the efficiency of our training strategy.

Instead of training a new code language model (LM) from the ground up,We also compared different self-refining techniques and ours on the state-of-the-art models.ed a vast amount of information about code structure, syntax, and common patterns.

\sln{} then builds upon this foundation by training the model to self-refine its code generation. This focused approach allows \sln{} to enhance the self-correction capabilities of existing code LMs without the need for extensive and time-consuming training from scratch.

We also compared between different self refining techniques and ours on state of the art models. Namely GPT-4o and Gemini 1.5 002. As we can see in Table \ref{tab:gemini}.

\begin{table}[]
\begin{tabular}{|l|c|c|}
\hline
\textbf{Model} & \textbf{Approach} & \textbf{Move} \\ \hline
GPT-4o & Direct            & 27.8           \\ 
        & COT               & 32.6           \\ 
        & Self-Planning     & 30.2           \\ 
        & Self-Debugging (+Trace) & 29.0           \\ 
        & \textbf{ConMover(ours)} & \textbf{37.6}  \\ \hline
Gemini 1.5 002 & Direct            & 37.8           \\ 
               & COT               & 39.6           \\ 
               & Self-Planning     & 39.0           \\ 
               & Self-Debugging (+Trace) & 38.4           \\ 
               & \textbf{ConMover(ours)} & \textbf{56.9}  \\ \hline
\end{tabular}

\vspace{0.5cm} % Adds vertical space 

\caption{Results with different approaches and improvement against direct baseline approach. COT\cite{wei2022chain}, Self-Planning\cite{jiang2023self},Self Debugging\cite{chen2023teaching}.}
\label{tab:gemini}
\end{table}

We conducted evaluations using the Pass@1 metric, In
the set k iterations (10 in our experiment), the problem is
considered solved as long as it is solved successfully once.
We primarily tested the currently most outperforming GPT-4o
model and the Gemini 1.5 002 model.
\section{Discussion}

Our experimental evaluation shows that \sln{} performs better than existing other methods. This leads us to a discussion regarding \sln{}'s limitations and achievements.

\subsection{RQ1: Code Generation Based on Concept}

Our initial goal was ambitious: generate compilable Move code directly from natural language descriptions, circumventing the need for extensive datasets and costly pre-training.  The success of \sln{} \textbf{demonstrates the feasibility of this approach, showcasing the potential of leveraging a limited set of documentation and code examples to extract meaningful concepts for Move code generation}.

Interestingly, our experiments revealed that these concepts are most effectively utilized not as direct context for code generation, but rather as a mechanism for feedback, planning, and iterative refinement. By employing the extracted concepts to analyze and critique initially generated code, we observed significant improvements in accuracy and overall code quality. This led us to design a system that optimizes efficiency by encoding our knowledge base within a knowledge graph, dynamically queried based on the initial (potentially faulty) code to provide targeted feedback and guide the generation process. This approach minimizes reliance on large context windows and reduces token usage, contributing to a more streamlined and efficient code generation workflow.

\subsection{RQ2: Code Generation Capability use smaller data}

A key aspect of our approach was the deliberate limitation of our knowledge base and retrieval-augmented generation (RAG) pipeline. We utilized a concise dataset comprising only 20 Move smart contract categories sourced from the official Move GitHub repository \cite{suiexamplesmoveatmainMystenLabssui}. Even with this restricted dataset, we achieved promising code generation results.

Similar to our observation with concept utilization, this limited codebase proved most valuable in enhancing the planning agent's capabilities. This finding suggests a promising avenue for future research: expanding the knowledge base with additional Move code examples could further bolster the robustness and effectiveness of the planning agent, leading to even more accurate and reliable code generation.

\subsection{RQ3: Code Generation Capability for Small LLMs}

Our results demonstrate a consistent performance boost across all tested models when integrated with \sln{}, with larger models generally exhibiting greater improvement.  However, a particularly striking observation emerged from the Gemma-2 2B model.  Despite its significantly smaller size, it achieved performance nearly on par with its larger counterparts, Gemma-2 9B and CodeLlama. This suggests that, under certain conditions, smaller LLMs can generate code of comparable quality.

This outcome was particularly unexpected for Move, as smaller LLMs are typically not expected to possess the reasoning capabilities necessary to fully leverage the planning recipe provided by \sln{}. We hypothesize that this surprising performance stems from the unique training methodology employed for \sln{}. By exposing the model to both faulty and correct code during training, coupled with an active feedback mechanism focused on rectification, \sln{} develops a nuanced understanding of code quality. This allows it to not only distinguish between good and bad code but also to actively repair and refine code towards a correct solution. This inherent corrective capability provides \sln{} with a distinct advantage, enabling even smaller LLMs to generate high-quality Move code.

So we can say While maintaining the one-time code generation capacity, mostly improving
marginally, \sln{} significantly boosts the code LMs’ self-refine capacity by learning to understand the execution feedback and faulty code generated in the past. \sln{} enables existing
code LMs to match or beat larger models with 3× more parameters.

\subsection{RQ4: Self Correction for faulty code}
We believe the impressive performance of \sln{} is largely attributable to its robust feedback and debugging mechanism, driven by the planning agent. While we cannot definitively assert that \sln{} explicitly "repairs" code in the traditional sense, our empirical observations strongly support this notion. By iteratively feeding the \sln{} pipeline with faulty code augmented with contextual information, including potential execution traces and compilation feedback, we witnessed significant improvements in code generation accuracy compared to previous work by Karanjai et al. \cite{karanjai2024solmover} and Tarassow et al. \cite{tarassow2023potential}. This iterative refinement process, guided by the planning agent's insights, enables \sln{} to effectively learn from its mistakes and progressively generate higher-quality code.

\section{Conclusion}

This study highlights the importance of enabling open-source LLMs to self-debug and introduces a scalable framework to achieve this. Our framework integrates automated data collection, validation, supervised fine-tuning, and reinforcement learning with novel reward mechanisms to enhance the self-debugging capabilities of LLMs. Importantly, our data collection process operates independently of the LLM itself, allowing us to correct and improve the performance even with low-resource code generation tasks, such as generating Move code through a planning agent empowered by documentation and limited code examples.

We demonstrate the feasibility of generating code based on conceptual understanding using an agentic approach and a self-correction pipeline. Our proposed framework, \sln{}, offers a scalable and cost-effective solution to improve the code generation abilities of any off-the-shelf open-source LLM, eliminating the need for expensive training or massive datasets.

% \begin{credits}
% \subsubsection{\ackname} A bold run-in heading in small font size at the end of the paper is
% used for general acknowledgments, for example: This study was funded
% \end{credits}
\bibliographystyle{splncs04}
\bibliography{ref}
\end{document}